\documentclass[12pt]{article}
\usepackage{graphicx}
\usepackage{amsmath}

\newcommand{\beq}{\begin{equation}}
\newcommand{\eeq}{\end{equation}}

\newcommand{\hi}{h^{-1}}

\begin{document}

%\twocolumn[%
\begin{center}
{\bf \LARGE Fractal Holography: a geometric re-interpretation
of cosmological large scale structure}

%{\bf \large Is the large scale fractal distribution of galaxies a geometric
%signature of holography?}
\vskip .4cm
J. R. Mureika \\
{\small \it Department of Physics, Loyola Marymount University, Los Angeles, CA  90045-8227 \\
Email: jmureika@lmu.edu}
\end{center}
%\vskip 2cm

{\noindent{\bf Abstract} The fractal dimension of large-scale galaxy clustering 
has been reported to be roughly $D_F \sim 2$ from a wide range of redshift 
surveys.   If correct, this statistic is of interest for two main reasons: fractal scaling 
is an implicit representation of information content, and also the value 
itself is a geometric signature of area.  It is proposed that a
fractal distribution of galaxies may thus be interpreted as a signature of 
holography (``fractal holography''), providing more support for current 
theories of holographic
cosmologies.  Implications for entropy bounds are addressed. In particular,
because of spatial scale invariance in the matter distribution, 
it is shown that violations of the spherical entropy bound can be removed.
This holographic condition instead becomes a rigid constraint on the nature
of the matter density and distribution in the universe.  Inclusion of a dark
matter distribution is also discussed, based on theoretical considerations
of possible universal $\Lambda$CDM density profiles.}\\


\vskip 1.5cm

%\pagebreak

\section{Introduction}
\label{intro}
The popular notions of fractals revolve around spatial power law scaling, 
physical self-similarity, and structural recursiveness \cite{mandel1}. 
Mathematically, this relationship assumes the general form
\begin{equation}
N(r) \sim r^{D_F}~~.
\label{fracdim}
\end{equation}
where $D_F$ is the fractal dimension and $r$ is the scale measure.  The
quantity $N(r)$ represents the characteristic of the distribution that 
exhibits the fractal behavior.  Measurement of fractal
statistics for a wide range of physical phenomena has been addressed over 
the years, ranging from the shape of coastlines to  the structure of clouds,
and pertinent to this paper, the large scale distribution
of visible matter in the universe \cite{peebles2}.  

It should be emphasized that the meaning of the fractal dimension is not 
only statistical, but it also
has geometric significance.  Topological considerations constrain the 
fractal dimension of a distribution to be less than (or equal) to that of the
space in which the structure is embedded \cite{falconer}.  Furthermore,
when a fractal dimension coincides with an integer dimension,
it is possible to make the association between the structure under 
consideration and the geometry associated with the dimension.  That is,
a distribution with fractal dimension of $D_F = 0$ is described as a 
point distribution,
$D_F = 1$ a linear distribution, $D_F = 2$ a surface distribution,
and $D_F = 3$ a volumetric or space-filling distribution.

A hierarchically-structured universe is a recurrent theme in our understanding 
of Nature.  Since at least the early sky maps of Charlier \cite{charlier}, it
has been suggested that galaxies do {\it not} cluster in a random fashion, but 
rather appear in clumps interspersed by voids.  The advent of deep sky redshift surveys brought with it a surge
interest surrounding the exact nature of large-scale galaxy
distributions in the observable universe.   An overwhelming number of 
independent estimates of the galaxy clustering fractal dimension, obtained 
from a variety of sources seem to unanimously suggest that this statistic 
has a value of or around $D_F = 2$.   Up to the release of the SDSS data,
the various redshift surveys had probed depths up to at least $10\hi~$Mpc and
confirmed the fractal scaling behavior.  Extrapolating the analysis to 
include superclustering structure suggested this behavior continued 
well up to $100-1000~\hi$~Mpc \cite{labini}, with no apparent transition to 
homogeneity. 

\begin{table}
\begin{center}
{\begin{tabular}{l c c c c c}\hline
Survey & $D_F$ & Approx.\ Size \\ \hline
CfA1 & 1.7 (0.2)& 1800 \\
CfA2 & $\sim 2$& 11000\\
SSRS1 & 2.0 (0.1)& 1700\\
SSRS2 & $\sim 2$& 3600 \\
LEDA & 2.1 (0.2)& 75000\\
IRAS 1.2/2~Jy & 2.2 (0.2)& 5000 \\
Perseus-Pisces & $\sim 2.1$&3300  \\
ESP & 1.8 (0.2) & 3600 \\
Las Campa\~{n}as (LCRS) & 2.2 (0.2)& 25000\\ 
SDSS (r1) & $\sim 2$ & $2\times 10^5 - 1.5\times 10^6$ \\ \hline
\end{tabular}}
\end{center}
\caption{
Galaxy fractal dimension calculations for various redshift surveys
(compiled from \cite{labini},\cite{baryshev},\cite{jaswant},\cite{labini05}).
}
\label{tab1}
\end{table}

The newest SDSS 
redshift data confirms the $D_F \sim 2$ to a high precision up to $20\hi$~Mpc, 
but with the correlation weakening to homogeneity at distances of $70\hi$~Mpc.
Alternate analyses suggest that the transition to homogeneity is not observed,
but instead the fractal scaling continues up to $200\hi~$Mpc \cite{baryshev}.
The authors of Reference~\cite{jaswant} perform a two-dimension multifractal
analysis on SDSS projected data, deducing that the appropriate scaling is
confirmed to dimensions $D_q \in (1.7,2.2)$ for all positive and negative $q$.

On the whole, this observational data suggests a (local) violation of the 
cosmological principle.  Geometrically speaking,
a homogeneous visible universe should manifest itself
as an $N(r) \sim r^3$ distribution.  In terms of the fractal dimension, this is
a volumetric scaling with a dimension $D_F = 3$, synonymous with the lack of 
any preferential direction.  The origins of the observed large 
scale structure in the universe are unknown, although it is commonly believed 
that it has arisen from anisotropically-distributed quantum fluctuations 
in the pre-inflation epoch.  The recent analysis of the SDSS 
data \cite{sdss2} coupled with emerging CMB data from WMAP \cite{wmap} data 
has helped to isolate probable cosmological parameters to determine 
viability of formation models.

Although the origin of this structure is still a mystery, the most successful models are
$\Lambda$CDM gravitational collapse scenarios in the early universe \cite{davis}.
The basic framework of such cluster-formation models relies on perturbations in 
the local background curvature due to quantum fluctuations during the inflation epoch.
Such models successfully reproduce a wide range  of observable clustering features,
including power spectra and correlation functions, as well in many cases
the cited fractal dimensions for galaxy clustering \cite{apj394,borgani,dubrelle,colombi}.
More recently, such models have been extended to galaxy and quasar clustering \cite{volker},
with the added advantage of introducing universal dark energy constraints.
In this sense, the $\Lambda$CDM scenarios are the model logical and consistent
explanation for the present inhomogeneous matter distributions.  In fact, very recent 
observational data \cite{massey} supports the filamentary dark matter ``scaffolding'' that
are a consequence of the models.

%A number of theoretical solutions have been offered for such inhomogeneous
%structure, based on CDM N-body gravitational collapse scenarios
%These are of particular interest due to their natural connection to
%hierarchical clustering growth from small initial mass/density perturbations
%in the early universe.  

Despite the fact that the apparent success of the $\Lambda$CDM models seems to put to rest
any mysteries surrounding the origin of large scale structure formation, this paper
will focus on a much different mechanism.  which can be explained by adopting
a new perspective on how matter and gravitation are allowed to behave?
Fractal behavior is generally not associated with equilibrium growth, and
thus most models of large scale structure evolution do not predict its
existence.  As the aforementioned evidence suggests, however
there appears to be a uniformly-defined fractal distribution of matter in the universe,
at the very least up to some as-yet unkown scale length.
The use of entropy to represent fractal structure stems
from the implicit relation between entropy and information (this is discussed
in the concluding section of this paper).   Fractal -- and moreover multifractal
-- statistics quantify the nature in which information is encoded or distributed
in a system.  It is the intention of this paper to highlight this connection 
between information, entropy, and fractality.

Before proceeding, it would be negligent to ignore the controversy surrounding the
fractal interpretation of large scale structure, which is by no means an established fact.  
Dubbed the ``fractal debate'', there has
been an ongoing discussion of whether or not the analyses indicating fractal clustering
have been performed with the proper data treatment, most notably a reliable estimate
of the redshift distance from otherwise two-dimensional projections.   There are clearly
two opposing opinions on the exact nature of any characteristic scale lengths and clustering
that might exist in clustering.  The interested reader is directed to some comprehensive
and competing summaries of both sides of the debate in such references as 
\cite{labini,eso,martinez2,rees,abdalla2}.
The availability of more redshift data will either confirm, deny, or further muddle this
issue.  For the purposes of this discussion, however, the fractal model will be assumed to
be correct.

More pertinent to this paper, it should be noted that fractals themselves are
members of a fundamental class of statistical object whose basis lies in the 
heart of information theory.  A q-multifractal is defined by the measure 
partition $Z(q,r) = \sum_i [p_i(r)]^q$, and dimensions 
$D_q = \left(\frac{1}{q-1}\right) \; \frac{d \log[Z(q,r)]}{d log [r]}$.
where $q$ is any integer and $p_i$ is the local spatial density of the 
fractal object within a sphere of radius $r$ \cite{falconer}.  
The traditional 
fractal dimension is obtained in the limit $q = 0$, but when $q \rightarrow 1$ 
this quantity becomes $I(r) = -\sum_i p_i(r)\; 
\log p_i(r)~,~D_1 = \frac{d \log I(r)}{d \log r}$.
The function $I(r)$ is the Shannon information entropy \cite{ref2}.  
In the case of a monofractal distribution, all dimensions ${D_q}$ collapse to
the single value $D_0$, which is the fractal dimension.  In this respect,
the fractal dimension may be seen as a representation of a system's entropy
measure and content.

\section{Information theory and the holographic principle}
Information content and entropy have entered the area 
of the long-standing dichotomy between classical and quantum gravity.
In early works by Beckenstein
\cite{beckenstein} it was suggested that the maximum entropy contained within a
black hole was determined not by its volume, but rather the horizon area
${\cal A}_H$.  This limit, known as the Beckenstein bound, placed stringent
constraints on how information could distribute itself within a region
of space.  It was further generalized as the spherical entropy bound,
\begin{equation}
S(V) \leq \frac{A}{4}
\label{hp1}
\end{equation}
where $S(V)$ is the entropy contained in a volume of space $V$, and
$A$ is the area of the (spacelike) boundary of $V$
(in units of the Planck area).

Bousso has shown that each of these entropy bounds can be understood as
classes of a more general theory known as the holographic principle (HP)
\cite{bousso}.  An unproven hypothesis, the HP suggests that there exists
a deeper geometric origin for the total number of possible quantum states
which can occupy a spatial region.  In its most general formulation, the
HP states that $S(B) \leq \partial B /4$, where $B$ is some region, 
$\partial B$ its boundary, and $S(B)$ the entropy contained in $B$.
Although most instances are subject to specific failures,
the most radical formulation of the HP -- Bousso's Covariant Entropy Conjecture
-- proposes that the entropy bounded by a region $B$ defined by the 
{\it light sheets} of a backward-pointing null cone obeys a holographic-type
relationship.
The various entropy bounds which form the holographic principle place rigid 
constraints on the number of possible entropy states which can occupy a 
region of space \cite{bousso}.  

The HP is novel in its motivations: the physics of a spatial $n$-dimensional
region are defined by dynamical systems which exist exclusive on the 
region's (n-1)-dimensional boundary.  The most promising support of this 
theory is the AdS/CFT correspondence \cite{adscft}, which explicitly connects 
via a one-to-one correspondence the framework of a 5D string theory in 
anti-deSitter space with a conformal quantum field theory on the 4D boundary.

\section{Cosmology and holography}
\label{cosmicholo}
Fischler {\it et al.} \cite{fisch1,fisch2,fisch3,fisch4}
have proposed an extensive cosmological version 
of the theory, primarily based in part on the spherical entropy bound.
Dubbed ``Holographic
Cosmology'', the framework presents an alternate inflationary 
evolutionary model for large scale structure in which structure evolves from evaporating
primordial quantum black holes.  This model not only matches current
observation but also explains the flatness and horizon problems.  A similar
proposal based on the original work of Susskind and Fischler is discussed
in \cite{bak}, which proposes a ``cosmic holography'' bound in FRW universes
of positive, flat, and negative curvature.  Further comparisons and contrasts
between holography and cosmology are offered in \cite{kaloperlinde}, in which
constraints from inflation are the focus.
Similarly, the authors of \cite{abdalla3} also
show that power spectrum correlations and suppression in the CMB may
by holographic in origin.  References~\cite{abdalla4} discuss holographic
implications for (2+1)-dimensional cosmological models.

As derived in \cite{fisch1},
assuming a homogeneous and isotropic universe with 
constant mean density, it is possible to define a (co-moving) entropy
density $\sigma$ such that the total entropy with a volume $V$ is \cite{bousso}
\begin{equation}
S  = \int_V \; d^3x \sqrt{h}\sigma
\end{equation}
which can be written $S = \sigma V$, where $\sigma$ is the (volumetric)
entropy density.  The entropy condition is thus
\begin{equation}
\sigma V \leq \frac{A(V)}{4}~~,
\label{seb}
\end{equation}
where $A(V) = 4 \pi r^2$ is the bounding area of the volume $V$ (in flat
space).  Including the $r$-dependence, the entropy bound is
\begin{equation}
r^3 \sigma \leq \frac{3r^2}{4}~~,
\label{bousso1}
\end{equation}
and so it can easily be shown \cite{bousso} that the spacelike entropy bound is violated
for sufficiently large values of $r > 3/4\sigma$.

%NEW
The aforementioned models rely on quantum mechanical entities (black holes) to describe
the basic units of the entropy bound, while proposals like the one presented herein
assume that galaxies are the key.   The application of holographic bounds
makes the assumption that these objects are themselves in some way a fundamental ``unit''
of entropy.  It is, however, reasonable to make this {\it a priori} postulate.
% for several reasons.
%Classical limits of quantum theories are quite widespread.  Ehrenfest's theorem, for example,
%cleanly relates the time-evolution of a system's quantum mechanical expectation values to the appropriate deterministic classical behavior.    A similar quantum/classical connection analogy can be made for the correspondence principle.  The analogies are very limited, though, since ultimately the quantum and classical descriptions are fundamentally different.   Cosmological holography shares the same form as its quantum analog.

%NEW
It has long been theoreized that galaxies themselves
are the result of gravitational clustering around supermassive black holes that have either grown from accretion, or from combining with smaller primordial black holes (see the various references \cite{gbh1,gbh2} and references therein).  Although supermassive black holes had been thought only to populate the bulges of active galactic nuclei, growing evidence suggests that such objects may also be found in type I Seyfert galaxies  \cite{ghb3} and high-redshift blazars \cite{ghb4}.  If one considers black holes to be the ``seeds'' of every galaxy, then these objects may be understood to represent a universal evolution toward a maximal spatial entropy state, independent of the type or size of galaxy.

\section{A fractal connection to holography?}

The work presented herein is similar in inspiration to that of Fischler {\it
et al.}, and like those referenced works promotes the notion that holography
should be a viable candidate for a constraint of structure evolution.  
Holographic constraints will first be applied to visible matter, but a discussion 
including dark matter distributions will follow in Section~\ref{darkmatter}.  
Since it is somewhat different from the
previous cosmological holographies proposed in the literature, it might
be appropriate to label this version as ``fractal holography''.

It is the different spatial dependences of area and volume that allows 
the inequality
(\ref{seb}) to be violated.  With respect to Equation~\ref{fracdim}, however,
one can reformulate the ``problem''
of large scale fractal clustering by focusing not on 
the apparent break from homogeneity and isotropic scaling, but rather
by highlighting the specific {\it geometry} of the scaling.  
Since the redshift surveys cited in Table~\ref{tab1} indicate that 
large-scale matter is distributed according to a $D_F=2$ scaling, it seems 
more appropriate to describe the entropic content by the mean ``surface'' 
entropy density $\xi$.  That is, each object contributes an average
entropy $S(V) / N$ and there are $N \sim r^{D_F}$ objects.  
It should be noted that the fractal power laws are derived from average density
considerations, so such an argument is certainly well-founded.
Fractal large scale structure thus states that within a sphere of radius
$r$, the number of cosmological objects is a function of {\it area} ($r^2$). 

Thus, within a spherical volume of radius $r$, the number of 
galaxies $N(r)$ must be {\it proportional} to the surface area of the 
region's boundary,
\begin{equation}
N(r) \propto \partial V(r) = A(r)~,
\label{areascale}
\end{equation}
so that the entropy contained with a region $V$ is
\begin{equation}
S(V) =\alpha \xi A~~,
\label{areaent}
\end{equation}  
where $\alpha > 0$ is the proportionality constant.
The above relation suggests that the
distribution of matter in the universe has perhaps a more fundamental and
geometric origin.  

In this case, a holographic-type space-like entropy bound is precisely given as 
\begin{equation}
S(V) = \bar{\xi} A \leq \frac{A}{4}~~,
\label{fracholo}
\end{equation}
where for simplicity the proportionality constant has been absorbed into 
the surface density term, $\bar{\xi} = \alpha \xi$.
Due to the $A \sim r^2$ dependence of each component of this inequality, the
spatial dependence vanishes and what is left is a truly scale-invariant
bound.  Specifically, the violation of the space-like entropy
bound is eliminated, and instead is replaced by rigid constraints on the
surface entropy density, and hence the {\it geometry} of the matter
distribution: 
\begin{equation}
\bar{\xi} \leq \frac{1}{4}~.
\end{equation}

What might be the value of $\xi$?
The fractal distribution of galaxies extends to about 10~Mpc, or
$10^{58}$ Planck length units.  The area of the bounding sphere is thus on
the order of $10^{116}$ area units.  The entropy content of the {\it entire} 
visible 
universe is on the order of $10^{90}$ \cite{kaloperlinde}, so even if 
a sizable fraction is represented in this fractal distribution, this implies
the ``surface'' density is no greater than $\xi \sim 10^{-24}$ or so.
The value of the proportionality constant $\alpha$ thus is the key to the
inequality.  Unless $\alpha$ is of exceedingly high order of magnitude, though,
it is unlikely that this bound will ever be violated.

\subsection{Entropy bounds for fractal dimensions near 2}
Although the observational evidence points to a fractal scaling dimensions
of $D_F = 2$, this exact geometric signature could be a coincidence.  If such
is the case, then the spherical entropy bound (\ref{fracholo}) will not
be scale invariant.  However, implications of the bound become even more
interesting if one follows the prescription for non-integer scaling dimensions
around $D_F =2$.

Following this philosophy, (\ref{fracholo}) can be expressed as
\begin{eqnarray}
\chi r^{D_F}&\leq&\frac{A}{4} \nonumber \\
\frac{\chi}{\pi} &\leq&r^{2-D_F}
\end{eqnarray}
where $\chi$ is the ``fractal number density'' of the distribution.
So, violations of the entropy bound will occur whenever (up to geometric
factors)
\begin{equation}
r > \left(\frac{1}{\chi}\right)^{1/(D_F-2)}~~.
\end{equation}

If the fractal dimension is slightly higher than 2, the bound will ultimately be
violated for a large enough sphere.  However, the relative radius of the
sphere will be much greater than in the case of homogeneity.  
In the case $D_F=3$, this reduces to the violation derived in \cite{fisch1}.

\subsection{Transitions to homogeneity}
\label{homtrans}
Current observation suggests that the fractal distribution of matter
may transition to homogeneity at large distances.  In this case, the
entropy density (\ref{areaent}) and scale-invariant entropy bound 
(\ref{fracholo}) are no longer applicable, at least on a global scale.

In addition to the references discussed in Section~\ref{intro},
a recent analysis \cite{hogg} has determined
that number of luminous red galaxies (LRGs) shows a well-defined
$D_F = 2$ fractal behavior up to scale lengths of at least $20\; h^{-1}~$Mpc,
and a smooth
transition to homogeneity ($D_F = 3$) beyond scales of $70\; h^{-1}~$Mpc.
Reference \cite{vasilyev} confirms $D_F \sim 2$ fractal clustering behavior
to a scale of $40~$Mpc, using independent analysis techniques such as the
nearest neighbor probability density, the conditional density, and the
reduced two-point correlation function.

In terms of the holographic model, in the simplest of cases, consider a spherical region of radius $R$
(in flat space) in which the distribution of matter is fractal with
$D_F = 2$.  Within this region, the entropy constrain obeys that 
described in Equation~\ref{areaent}, 
{\it i.e.} $S_{\rm F}(r) \sim \bar{\xi} r^D_F$.  
For separation scales $r > R$,
the distribution resembles a homogeneous one, and the total
entropy is now described by a volumetric distribution with 
density $\sigma$.  At the transition scale $r=R$, these descriptions of
entropy must agree.  That is, the total entropy within a sphere of
radius $r=R$ should be $S_{\rm F}(R) = S_{\rm H}(R)$, where 
$S_{\rm F}(R) =4\pi \bar{\xi} R^2$ and $S_{\rm H} = 4/3 \pi \sigma R^3$.
This implies that $R = \frac{3\bar{\xi}}{\sigma}$.

Potential violations of the holographic principle are now re-introduced
at scales $r>R$, as described by Equation~\ref{bousso1}.  However, the
requirement of statistical continuity in the description of the entropy
also requires that $S_{\rm F}^\prime(R) = S^\prime_{\rm H}(R)$.  This
provides an additional constraint equation on the value of $R$, in this
case $R = \frac{2\bar{\xi}}{\sigma}$, assuming the only radial dependence
in the holographic bound is in the geometric term (area and volume).  

%places strong constraints on the nature of the matter distributions at
%scales $r < R$.  Specifically, this implies $\bar{\xi} = \sigma R$.
%Therefore, the combination of the total entropy in the universe and the
%homogeneity scale $R$ explicitly determines the ``density'' of matter on 
%smaller scales.   This gives new interpretation of $R$ as a type of
%critical parameter that marks some variety of  ``phase transition'' in the
%distribution of matter.
%The exact ``location'' of the transition is currently unknown.  

Hence, fractal holography implicitly supports the well-known observation of a
slow transition to homogeneity over distances of several megaparsecs.  A similar
holographic model could represent the transition entropy in the most general 
form $S_{\rm Trans}(r) \sim \rho(r)r^{D(r)}$, where both the entropy ``density'' ($\rho(r)$) and
the fractal dimension itself ($D(r)$) become functions of the scale length.
This can be cast as a boundary-constrained problem, with the requirements
$S_{\rm F}(R_1) = S_{\rm Trans}(R_1)$, $S_{\rm H}(R_2) = S_{\rm Trans}(R_2)$,
$S^\prime_{\rm F}(R_1) = S^\prime_{\rm Trans}(R_1)$, 
$S^\prime_{\rm H}(R_2) = S^\prime_{\rm Trans}(R_2)$.  Appropriate constraints on
the values of $D(r)$ and its derivative, for example, could be isolated from correlation
and fractal analyses.

It should also be noted that a critical discussion of 
the meaning of ``transition to homogeneity'' found in \cite{gatte} indicates 
that there could be two possible
interpretations of what it means to transition to homogeneity.  This could
be in terms
of a trivial correlation function at $r=R$ (the usual transition boundary), as
well as a more long-range scale $\lambda$ that could be greater than the
horizon distance.  Regions measured at scales $R < r < \lambda$ are not
strictly homogeneous, but rather are analogous to a fluid at the critical point.

An oft-cited argument for the necessity of a transition to homogeneity stems from the exceedingly homogeneous distribution of anisotropies in the cosmic microwave background (CMB).   It is important to note, however, that inferring homogeneity of a three-dimensional distribution from its two-dimensional projection is not trivial.  From the point of view of fractal statistics, for example, a homogeneous surface distribution would coincide with a fractal dimension of $D_S=2$.    The fractal projection theorem \cite{vicsek} states that for any fractal with dimension $D_F$, its projection onto a sub-plane of dimension $D_P \leq D_F$ will itself have a dimension of $D_P$.  That is, the dimensionality of the projective distribution ``saturates'' the sub-plane.   It is thus possible that the perceived homogeneity of the CMB may not truly correlate to the homogeneity of volumetric spatial distributions at the time of last scattering.  In this case, the original distribution of anisotropies may well have obeyed a holographic-type ``area'' constraint, like that discussed herein.  

\subsection{Scale evolution and fractal holography}
There is ongoing debate as to whether or not the fractal distribution
of visible matter extends indefinitely to beyond 1000~$\hi~$Mpc.  If
in fact the entire universe is governed by the $D_F \sim 2$ fractal 
distribution, the fractal holographic condition can place a bound on 
its expansion rate.

For a sphere whose radius $R_H$ is the horizon distance, the usual
holographic bound in $d$ spatial dimensions is \cite{fisch1}
\begin{equation}
\sigma R_H(t)^d < [a(t) R_H(t)]^{d-1}~~.
\end{equation}
with $a(t)\sim t^p$ the scale factor of the universe, $p$ an expansion
parameter, and 
$R_H(t) = \int_0^t a(t^\prime) dt \sim t^{1-p}$.
The left hand term in the inequality assumes
that the entropy scales volumetrically.  Since $\sigma$ is small,
it has been demonstrated that the inequality is satisfied throughout
the history of the universe if $p > 1/d$.

Adopting the fractal interpretation and setting $d=3$, this becomes
\begin{equation}
\bar{\xi} R_H(t)^2 < [a(t) R_H(t)]^2~~.
\end{equation}
The new constraint on the evolution parameter is thus $\bar{\xi} < t^{2p}$,
which is almost certainly always satisfied for an arbitrary choice of $p$.

\section{Inclusion of dark matter}
\label{darkmatter}
So far, the discussion of fractal holography has excluded dark matter.  
Clearly, any viable model for large scale structure must replicate more
than the visible matter distributions, but also the ``invisible'' ones.
Although the spatial structure of halo dark matter density profiles can 
easily be inferred from galaxy rotation curves, it is
uncertain exactly what the large scale distribution looks like.  

Since dark matter is believed to make up well over 90\% of the material
content of the universe, no cosmological model would be complete without
paying due attention to this mystery.  Unfortunately, not much is known
about the actual form of the distribution of dark matter in the universe.
The best models we have for density distributions are those of ``small scale''
dark matter halo structures derived from galaxy rotation curves, such as
the NFW profile \cite{nfw}, which suggest
density profiles of the form $\rho_{\rm DM} \sim r^{-2}$.   While these reflect the 
distribution of halo dark matter, they unfortunately offer no insight into the larger
scale structure.

Based on a simple inverse-square density profile for dark matter, 
the authors of Reference~\cite{durrer} have shown that a fractal distribution of galaxies
is not inconsistent with a homogeneous distribution of all matter, by virtue
of the fact that the dark matter density profile is the functional 
reciprocal of the galaxy number count.  The authors further note that this
implies a different fractal correlations for luminous matter ($D_F = 2$) and
dark matter ($D_F = 3$).  

Some numerical simulations of $\Lambda$CDM cosmologies suggest that 
dark matter halo profiles should roughly echo that of the matter distribution
in the universe \cite{livingreview}.   In particular, a universal density profile derived
from N-body simulations has shown that dark matter may cluster in a hierarchical
fashion \cite{navarro}.  

It has more recently been suggested that all baryonic matter can be distributed
in an  $r^2$ fractal manner, by appealing to alternative theories such as Modified Field
Theory \cite{kirillov}.  Such a description is consistent with both the Cosmological
Principle, as well as the Silk Effect \cite{silk}, and can produce a gravitationally-stable 
fractal clustering (the interested reader is referred to \cite{kirillov}
for further details).

For simplicity, assume the density profile of dark matter is hierarchical, according
to the power law $\rho_{\rm DM} \sim r^{-\gamma}$.  This implies that the number
of objects within a region of radius $r$ is $N_{\rm DM} \sim r^{3-\gamma}$, which
we may associated with a fractal dimension $D_{\rm DM} = 3-\gamma$.  The 
value of $\gamma$ ranges depending on the literature source, from $\gamma = 1.5$
\cite{dmhalo} to $\gamma = 2.1-2.5$, thus the corresponding fractal dimensions would
range between $D_{\rm DM} \sim 1.5-2.5$.  In this case, if $D_{\rm DM} \leq 2$, the
holographic constraint behaves as with luminous matter (and thus is potentially
not violated).

The most promising glimpse at possible larger-scale distributions of dark matter has been
reported  by the COSMOS collaboration \cite{massey}.  These results suggest a largely filamentary clustering of dark matter in ``rods'', with the more familiar halo clusters forming at ``vertices'' of several rods.  The extact spatial extent of the observed filaments is difficult to extract from the present data, and specific clustering measurements are currently underway\footnote{R.\ Massey, personal communication}.   Preliminary results suggest that the dark matter filaments can extend up to 30~Mpc or more \cite{massey}.  Nevertheless, the global geometry of such inhomogeneous filamentary structures would be consistently with a linear dark matter scaling (similar to a fractal scaling of $N(r) \sim r$), which would not violate any holographic constraints.

\subsection{General density distributions and holographic charge}
As a final comment on the geometric re-interpretation of the holographic principle, consider
the general case of matter distributions with fractal dimensions $D_F = 0, 1, 2, $and $3$.
The holographic constraints are thus dimensionally written (omitting some constants)

\begin{eqnarray}
S_0~ , ~S_1 ~, ~S_2 ~, ~S_3~\leq~ \frac{A}{4} \nonumber \\
\delta ~, ~\lambda r ~, ~\xi r^2 ~,~ \sigma r^3 ~\sim ~r^2 \nonumber \\
\frac{\delta}{r^2} ~,~ \frac{\lambda}{r} ~,~\xi ~,~\sigma r~ \sim~ 1 
\label{entropyfield}
\end{eqnarray}

The above expressions, cast in this manner, now become geometrically reminiscent of
field strengths for various charge distributions.  The geometry in
question corresponds to the dimensionality of the
fractal: point, linear, surface, and interior field of a continuous distribution.    The
functional similarity to Gauss' Law comes from the area term of the inequality, and
in this sense one might interpret Equation~\ref{entropyfield} as some variety of
``holographic field'' equation (although the similarities are most likely only superficial).

\section{Conclusions and future directions}
The inspiration for cosmological holography rests in the notion that quantum 
scale entropic physics can be arbitrarily expanded to any stable gravitational 
system.  It is thus possible that such holographic constraints in the early 
universe led to the formation and non-homogeneous distribution of anisotropies. 
Furthermore, the previous discussion suggests that the link between holographic 
area bounds and fractal $D_F = 2$ scaling may be related.  The galaxy 
number counts can be directly related to the entropy content by cosmological 
considerations like those discussed herein.

How is such a theory useful to further understanding cosmology and the origins of large scale structure?  At the very least, it helps to place somewhat rigid constraints on the nature of the clustering, as discussed in Section~\ref{homtrans} -- that is, where the clustering can be represented as ``fractal'', where it might be homogeneous, and over how far a distance the transition can occur.  Hence, a cosmological model can be built using the holographic constraints placed on the density distributions and their associated spatial extents.  In this sense, such a theory is not limited to only the ``fractal'' region of galaxy distributions, but can extend indefinitely and still provide useful boundary constraints.

Nothing in this proposal changes the fundamental origins of clustering.  Galaxies can still form as standard theory dictates.  Per the discussion in Section~\ref{cosmicholo}, galaxies do not replace anything as the fundamental unit of structure, since they themselves are likely ``perturbations'' to optimal clustering around black holes.  Since holographic theories have evolved from the study of such objects, it is not out of the question to expect galaxies to abide by some similar form of such principles.

Granted, the success or failure of any holography-inspired theory hinges on the success of the holographic principle itself.  At present, this is still a largely-hypothetical and unproven conjecture with limited (but growing) theoretical support.   Since the holographic bounds are derived as a fundamental ``entropy saturation limit'' of the universe, violations thereof would likely imply that the base assumption is incorrect.  

This paper has not considered the holographic bounds introduced in open
and closed universes, but there are several reasons for this omission.  First,
the wealth of observational data suggests that the universe is most likely
flat.  Secondly, while it is still possible to measure fractal dimensions in 
curved spaces using geodesic radii (via methods such as correlation analyses),
the elegant geometric interpretation of the fractal dimension 
(Equation~\ref{fracdim}) is not as easily realized.

Thus the connection between information theory, gravitation, and 
geometry is a common ``theme'' for fractal large scale structure. 
At the very least, the observed fractal 
distribution behavior of galaxies could be understood to be a large scale 
bookend principle to holography.  Redshift survey results provide strong 
evidence that the number counts scale as an area, but in order to verify a 
deeper connection future analyses should also focus on the 
pre-factor of the fractal relationship.  Fractal clustering of 
large-scale structure may well represent either a  
manifestation of holographic entropy bounds, or the end result of a
cosmological holography model, and future studies should adopt such a 
re-interpretation to explore new implications of the data.

\vskip1cm
\noindent{\bf Acknowledgments}\\
Thanks to Richard Massey (Caltech) for some insight regarding the COSMOS collaboration data.  JRM is support by a Research Corporation Cottrell College Science Award Grant.

\end{document}